\documentstyle[] {article}
\newcommand{\beq}{\begin{equation}}
\newcommand{\eeq}{\end{equation}}

\begin{document}

\begin{center}
\large
{\bf Classical and Quantum Perturbation Theory\\ 
for two Non--Resonant Oscillators \\
with Quartic Interaction}\\
\vskip 1. truecm
\normalsize
{\bf Luca Salasnich}\\
\vskip 0.5 truecm
Dipartimento di Matematica Pura ed Applicata, \\
Universit\`a di Padova, Via Belzoni 7, I--35131 Padova, Italy\\
\vskip 0.3 truecm
Istituto Nazionale di Fisica Nucleare, Sezione di Padova,\\
Via Marzolo 8, I--35131 Padova, Italy\\
\vskip 0.3 truecm
Istituto Nazionale per la Fisica della Materia, Unit\`a di Milano, \\
Via Celoria 16, I--20133 Milano, Italy\\
\end{center}

\vskip 0.5 truecm

{\bf Abstract.} We study the classical and quantum perturbation 
theory for two non--resonant oscillators coupled by a nonlinear 
quartic interaction. 
In particular we analyze the question of quantum corrections 
to the torus quantization of the classical perturbation theory 
(semiclassical mechanics). 
We obtain up to the second order of perturbation theory an explicit 
analytical formula for the quantum energy levels, 
which is the semiclassical one plus quantum corrections. 
We compare the "exact" quantum levels obtained numerically to the 
semiclassical levels studying also the effects of quantum corrections. 

\vskip 0.3 truecm

  
\vskip 0.5 truecm

\section{Introduction}

Nowadays there is considerable renewed interest in the 
transition from classical mechanics to quantum mechanics, 
a powerful motivation behind that being the problem of the so--called 
quantum chaos [1--3]. 
An important aspect is represented by the semiclassical quantization 
formula of the (regular) energy levels for quasi--integrable systems 
[4--6], the so--called torus 
quantization, initiated by Einstein [7] and completed by Maslov [8]. 
\par
It has been recently shown [9,10] that, 
for perturbed non--resonant harmonic oscillators, 
the algorithm of classical perturbation theory 
can be used to formulate the quantum mechanical perturbation theory 
as the semiclassically quantized classical perturbation theory equipped 
with the quantum corrections in powers of $\hbar$ "correcting" 
the classical Hamiltonian that appears in the classical algorithm. 
In effect, one can explicitly calculate corrections to the 
Einstein--Brillouin--Keller (EBK) quantization of the classical tori [4]. 
For example, the quantum corrections of the one--dimensional $x^4$ perturbed 
harmonic oscillator have been studied in great details by Alvarez, Graffi 
and Silverstone [11]. 
\par
Examples of rather detailed studies of semiclassical approximations, 
their resummations and of the Birkkoff--Gustavson normal forms can be found 
in Ali and Wood [12] and Ali, Wood and Devitt [13]. 
Another example of a rather complete 
semiclassical analysis of a one--dimensional system, namely 
the quartic oscillator, has been published by Voros [14]. 
\par
The aim of this paper is to extend previous studies 
to a two--dimensional system, which is more interesting because 
it is nonintegrable and thus generic. 
The integrable systems are rather exceptional in the 
sense that they are typically isolated points in the functional space of 
Hamiltonians and their measure is zero in this space. 
If we randomly choose a system realized in nature, the probability is one 
that the system is nonintegrable [15]. 

\section{Classical Perturbation Theory}

The model is given by two non--resonant oscillators 
coupled by a nonlinear quartic interaction of strength $g$ (Pullen and 
Edmonds [16]): 
\beq
H={\omega_1 \over 2}(p_1^2+q_1^2)+{\omega_2 \over 2}(p_2^2+q_2^2)
+ g q_1^2q_2^2. 
\eeq
Note that the same Hamiltonian has been obtained for the 
Yang--Mills--Higgs classical mechanics. The non--integrability 
of this system and its transition from order to chaos has been 
studied in classical with Poincare sections and in quantum mechanics 
with the spacing distribution of energy levels [17,18].  
\par
Through the canonical transformation in 
action--angle variables [19,20]: 
\beq
q_k=\sqrt{2I_k}\cos{\theta_k}, \;\;\; p_k=\sqrt{2I_k}\sin{\theta_k}, 
\;\;\; k=1,2 
\eeq
the Hamiltonian can be written: 
\beq
H=H_0(I_1,I_2)+ g V(I_1,I_2,\theta_1,\theta_2),
\eeq
where: 
\beq
H_0(I_1,I_2)=\omega_1 I_1+ \omega_2 I_2,
\eeq
\beq
V(I_1,I_2,\theta_1,\theta_2)=4I_1I_2\cos^2{\theta_1}\cos^2{\theta_2}.
\eeq
\par
Following the classical perturbation theory [19,20], 
we search for a canonical transformation 
$(I_1,I_2,\theta_1,\theta_2) \to ({\tilde I}_1,{\tilde I}_2,
{\tilde \theta}_1,{\tilde \theta}_2)$ to obtain a new Hamiltonian that 
depends only on the new action variables up to the second order in a 
power series of $g$:
\beq
{\tilde H}({\tilde I}_1,{\tilde I}_2)=
{\tilde H}_0({\tilde I}_1,{\tilde I}_2)
+g {\tilde H}_1({\tilde I}_1,{\tilde I}_2)
+g^2 {\tilde H}_2({\tilde I}_1,{\tilde I}_2).
\eeq
\par
The generator $S$ of the canonical transformation is supposed to be capable 
of expansion as a power series in $g$ of the form:
\beq
S({\tilde I}_1,{\tilde I}_2,\theta_1,\theta_2)=
{\tilde I}_1\theta_1 + {\tilde I}_2\theta_2 
+g S_1({\tilde I}_1,{\tilde I}_2,\theta_1,\theta_2)
+g^2 S_2({\tilde I}_1,{\tilde I}_2,\theta_1,\theta_2), 
\eeq
and to satisfy the equations:
\beq
I_k={\partial S\over \partial \theta_k}={\tilde I}_k
+g {\partial S_1\over \partial \theta_k}
+g^2 {\partial S_2\over \partial \theta_k},
\eeq
\beq
{\tilde \theta}_k={\partial S\over \partial {\tilde I}_k}=\theta_k
+g {\partial S_1\over \partial {\tilde I}_k}
+g^2 {\partial S_2\over \partial {\tilde I}_k}. 
\eeq
From the Hamilton--Jacobi equation: 
\beq
H_0({\partial S\over \partial \theta_1},{\partial S\over \partial \theta_2})+
g V({\partial S\over \partial \theta_1},{\partial S\over \partial 
\theta_2}, \theta_1, \theta_2)=
{\tilde H}_0({\tilde I}_1,{\tilde I}_2)
+g {\tilde H}_1({\tilde I}_1,{\tilde I}_2)
+g^2 {\tilde H}_2({\tilde I}_1,{\tilde I}_2),
\eeq
we have a number of differential equations that result on equating the 
coefficients of the powers of $g$:
\beq
{\tilde H}_0({\tilde I}_1,{\tilde I}_2)=
H_0({\tilde I}_1,{\tilde I}_2)=
\omega_1{\tilde I}_1+\omega_2{\tilde I}_2,
\eeq
\beq
{\tilde H}_1({\tilde I}_1,{\tilde I}_2)=
(\omega_1{\partial S_1\over \partial \theta_1} +
\omega_2{\partial S_1\over \partial \theta_2} )
+ V({\tilde I}_1,{\tilde I}_2, \theta_1, \theta_2),
\eeq
\beq
{\tilde H}_2({\tilde I}_1,{\tilde I}_2)=
( \omega_1{\partial S_2\over \partial \theta_1} 
+\omega_2 {\partial S_2\over \partial \theta_2} )
+ ( {\partial V\over \partial I_1}{\partial S_1\over \partial \theta_1} 
+{\partial V\over \partial I_2}{\partial S_1\over \partial \theta_2} ).
\eeq
The unknown functions ${\tilde H}_1$, $S_1$, ${\tilde H}_2$ and $S_2$ 
may be determined by averaging over the time variation of the 
unperturbed motion. \\
At the first order in $g$ we obtain: 
\beq
{\tilde H}_1({\tilde I}_1,{\tilde I}_2)= 
{1\over 4\pi^2}\int_0^{2\pi}\int_0^{2\pi} d\theta_1 d\theta_2 
V({\tilde I}_1,{\tilde I}_2, \theta_1, \theta_2) 
={\tilde I}_1 {\tilde I}_2 ,
\eeq
and 
$$
S_1({\tilde I}_1,{\tilde I}_2,\theta_1,\theta_2)=
-{1\over 4}{\tilde I}_1 {\tilde I}_2 
[{2\over \omega_1}\sin{2\theta_1}+{2\over \omega_2}\sin{2\theta_2}+
$$
\beq
+{1\over \omega_1 - \omega_2}\sin{2(\theta_1 -\theta_2 )}+
{1\over \omega_1 + \omega_2}\sin{2(\theta_1 +\theta_2 )}].
\eeq
At the second order in $g$ we have: 
$$
{\tilde H}_2({\tilde I}_1,{\tilde I}_2)=
{1\over 4\pi^2}\int_0^{2\pi}\int_0^{2\pi} d\theta_1 d\theta_2 
( {\partial V\over \partial I_1}{\partial S_1\over \partial \theta_1} 
+{\partial V\over \partial I_1}{\partial S_1\over \partial \theta_2} )
$$
\beq
=-{1\over 8}{\tilde I}_1{\tilde I}_2
[4 ({ {\tilde I}_1\over \omega_2 }+{ {\tilde I}_2\over \omega_1 })
- {({\tilde I}_1-{\tilde I}_2)\over \omega_1 -\omega_2}
+ {({\tilde I}_1+{\tilde I}_2)\over \omega_1 +\omega_2}].
\eeq
\par
We observe that the integrable approximate Hamiltonian (6) could be 
obtained alternatively as the Birkhoff--Gustavson normal form, which is 
a purely algebraic method of calculating the action variables order by 
order for perturbed harmonic oscillators with polynomial perturbations 
[21]. The Hamiltonian (6) depends only on the actions
so that by an application of the EBK rule: 
\beq
{\tilde I}_{1}=(n_1+{1\over 2})\hbar, \;\;\;\; 
{\tilde I}_{2}=(n_2+{1\over 2})\hbar, 
\eeq 
to equations (11), (14) and (16), we obtain a semiclassical 
analytical formula of the energy levels. Obviously $\hbar$ is the 
Planck constant and $n_1, n_2$ are two integer quantum number. 
\par
Note that the torus quantization can be applied because the normal 
form obtained (equation 6) is integrable. 
The new action variables remain very close to their initial conditions.  
This is no longer true, in general, for higher dimensions (see [19,20]). 
In the next section we show how to connect our semiclassical formula 
with the usual quantum perturbation theory. 

\section{Quantum Perturbation Theory}

In quantum mechanics the generalized coordinates 
are operators which satisfy the usual 
commutation rules for Bosons $[{\hat q}_k,{\hat p}_l]={\hat q}_k{\hat p}_l -
{\hat p_l}{\hat q}_k = i\hbar \delta_{kl}$, 
with $k,l=1,2$. Introducing the creation and destruction operators:
\beq
{\hat a}_k={1\over \sqrt{2\hbar}}({\hat q}_k+i {\hat p}_k),
\;\;\;\;
{\hat a}_k^+ ={1\over \sqrt{2\hbar}}({\hat q}_k -i {\hat p}_k),
\eeq
the quantum Hamiltonian can be written:
\beq
{\hat H}={\hat H}_0 + g {\hat V},
\eeq
where:
\beq
{\hat H}_0=\hbar \omega_1 ({\hat a}_1^+ {\hat a}_1 +{1\over 2})+
\hbar \omega_2 ({\hat a}_2^+ {\hat a}_2 +{1\over 2}),
\eeq
\beq
{\hat V}= {\hbar^2 \over 4} ({\hat a}_1 +{\hat a}_1^+)^2 
({\hat a}_2 +{\hat a}_2^+)^2. 
\eeq
\par
If $|n_1 n_2>$ is the basis of the occupation numbers of the two 
harmonic oscillators, the matrix elements are: 
\beq
<n_{1}^{'}n_{2}^{'}|{\hat H}_0|n_{1}n_{2}>=\hbar [\omega_1 (n_{1}+{1\over 2})+
\omega_2 (n_{1}+{1\over 2})]
\delta_{n_{1}^{'}n_{1}} \delta_{n_{2}^{'}n_{2}} ,
\eeq
and:
$$
<n_{1}^{'}n_{2}^{'}|{\hat V}|n_{1}n_{2}>=
{\hbar^2 \over 4}
[\sqrt{n_{1}(n_{1}-1)} \delta_{n^{'}_{1}n_{1}-2}
+\sqrt{(n_{1}+1)(n_{1}+2)}\delta_{n^{'}_{1}n_{1}+2}+
(2n_{1}+1)\delta_{n^{'}_{1}n_{1}}]\times 
$$
\beq
\times[\sqrt{n_2 (n_2-1)}\delta_{n^{'}_2 n_2-2}+ \sqrt{(n_2+1)(n_2+2)}
\delta_{n^{'}_2 n_2+2}+ (2n_2+1)\delta_{n^{'}_2 n_2}] .
\eeq
The Rayleigh--Schr\"odinger perturbation theory [22] up to the 
second order gives us:
\beq
E(n_1\hbar ,n_2\hbar )=E_{0}(n_1\hbar ,n_2\hbar )+g E_{1}(n_1\hbar ,n_2\hbar )+
g^2 E_{2}(n_1\hbar ,n_2\hbar ) , 
\eeq
where: 
\beq
E_{0}(n_1\hbar ,n_2\hbar )=\hbar [\omega_1 (n_{1}+{1\over 2})+
\omega_2 (n_{1}+{1\over 2})],
\eeq
\beq
E_{1}(n_1\hbar ,n_2\hbar )=<n_{1}n_{2}|{\hat V}|n_{1}n_{2}>,
\eeq
\beq
E_{2}(n_1\hbar ,n_2\hbar )= 
\sum_{\stackrel{n_1^{'}n_2^{'}}{(n_1^{'},n_2^{'})\neq (n_1,n_2)}} 
{ |<n_{1}^{'}n_{2}^{'}|{\hat V}|n_{1}n_{2}>|^2 \over 
\hbar [\omega_1 (n_1 -n_1^{'}) + \omega_2 (n_2 -n_2^{'})] }.
\eeq
\par
We obtain immediately:
\beq
E_{1}(n_1\hbar ,n_2\hbar )=\hbar^2 
(n_1+{1\over 2})(n_2+{1\over 2}),
\eeq
and after some calculations:
$$
E_{2}(n_1\hbar ,n_2\hbar )= 
{\hbar^3\over 32}[
{n_1(n_1-1)n_2(n_2-1)\over \omega_1+\omega_2}-
{(n_1+1)(n_1+2)(n_2+1)(n_2+2)\over \omega_1+\omega_2}+
$$
$$
+{n_1(n_1-1)(n_2+1)(n_2+2)\over \omega_1-\omega_2}-
{(n_1+1)(n_1+2)n_2(n_2-1)\over \omega_1-\omega_2}+
$$
$$
+{n_1(n_1-1)(2n_2+1)^2\over \omega_1}-
{(n_1+1)(n_1+2)(2n_2+1)^2\over \omega_1}+
$$
\beq
+{(2n_1+1)^2n_2(n_2-1)\over \omega_2}-
{(2n_1+1)^2(n_2+1)(n_2+2)\over \omega_2}  ].
\eeq
The zero and first order quantum terms coincide with the semi-classical ones:
\beq
E_{0}(n_1\hbar ,n_2\hbar )={\tilde H}_{0}
((n_1+{1\over 2})\hbar ,(n_2+{1\over 2})\hbar ), 
\eeq
\beq
E_{1}(n_1\hbar ,n_2\hbar )={\tilde H}_{1}
((n_1+{1\over 2})\hbar ,(n_2+{1\over 2})\hbar ),
\eeq
and the second order quantum term can be written: 
\beq
E_{2}(n_1\hbar ,n_2\hbar )={\tilde H}_{2}
((n_1+{1\over 2})\hbar ,(n_2+{1\over 2})\hbar ) + \hbar^2 
Q_2((n_1+{1\over 2})\hbar ,(n_2+{1\over 2})\hbar ),
\eeq
where:
\beq
Q_2((n_1+{1\over 2})\hbar ,(n_2+{1\over 2})\hbar )= - {3\over 32}
[{(n_1 - n_2 )\hbar \over \omega_1 -\omega_2} + 
{(n_1 + n_2 +1)\hbar \over \omega_1 +\omega_2}].
\eeq
The quantum series rearranges directly 
into the classical canonical perturbation series plus quantum corrections 
proportional to successive powers of $\hbar$ [10]. 
\par
The term $\hbar^2 Q_2$ represents the quantum corrections 
to the EBK quantization up to the second order of perturbation theory. 
These quantum corrections depend linearly 
on quantum numbers. For a quite similar system, Robnik [21] 
obtained that the spectra differ only by an 
additive constant independent of the quantum numbers (actions),  
but he studied only the first order of perturbation theory. 
\par
To conclude this section we mention 
the general problem of quantization: Only 
the quantization of coordinate space or the quantization 
of linear canonical transformations of coordinate space 
yields the right quantum mechanics, whose results agree with experiments. 
This is exactly the approach implemented in 
our present case (18), and thus our quantization is equivalent 
to the coordinate space quantization.  

\section{Numerical Results}

We compute the energy levels with a numerical diagonalization 
of the truncated matrix of the Hamiltonian (19) 
in the basis of the unperturbed oscillators 
(see [23] for a more technical discussion). 
The numerical energy levels depend on the dimension of the truncated matrix: 
We compute the numerical levels in double precision 
increasing the matrix dimension until the first 100 levels converge 
within $8$ digits (matrix dimension $1225\times 1225$). 
This means that the agreement between numerical values and real ones 
is very good and we use the word "exact" to indicate the numerically computed 
energy levels. 
\par
In Table 1 we compare the "exact" levels to the levels 
of the quantum perturbation theory and to the semiclassical results. 
A very good agreement is observed for the 
lowest energy levels. 
\par
In table 2 we show the error in units of the mean level spacing $D$ 
between the "exact" levels and the levels obtained with semiclassical 
and quantum perturbation theory. 
We observe that the algorithm provided 
by the appropriate semiclassical quantization is comparable to the algorithm 
provided by ordinary quantum perturbation theory but 
the quantal corrections do not always increase the accuracy. 
Thus for some of the calculated levels the semiclassical quantization 
gives better results than the quantum perturbation theory. 
\par
In Table 3 we show the error, in units of the mean level spacing $D$, 
between the "exact" levels and the semiclassical levels. 
As is seen, by decreasing $\hbar$ from $1$ to $10^{-1}$ the quality of the 
approximation improves considerably. However, it is important to 
observe that if $\hbar$, no matter how small, is kept fixed, 
the EBK quantization 
(torus quantization) of the individual levels is only a first order 
approximation of an expansion in $\hbar$. 
Therefore, {\it in general} 
the accuracy of the approximation decreases for higher levels. 
To get a good agreement it is necessary, as is well known, to implement 
the classical limit, i.e. $\hbar \to 0$ and $n_1,n_2 \to \infty$, 
while at the same time keeping the actions 
${\tilde I}_1 =(n_1 +1/2)\hbar$ and ${\tilde I}_2 =(n_2 +1/2)\hbar$ 
constant [23]. 

\section{Conclusions}

We have examined the transition between the classical 
and the quantum mechanics for a two--dimensional, nonintegrable 
and non--resonant system. 
Up to the second order of perturbation theory 
we have decomposed the quantum description into 
the classical description (i.e. the leading semiclassical term) 
plus quantum corrections which depend linearly on the quantum numbers. 
The semiclassical energy levels and 
the levels obtained with quantum perturbation theory 
are in good agreement with the "exact" numerical ones: 
The semiclassical quantization is comparable to the 
quantum perturbation theory, and for some levels the semiclassical 
quantization gives better results than quantum perturbation theory.  
\par
There are two series involved in the work. One in powers of $\hbar$ 
because of the semiclassical approximation and another in the interaction 
parameter $g$. The first one is an asymptotic expansion 
(see some very recent results in [24--27]) but the second one 
may be not convergent [28]. 
The classical and quantum perturbation series typically 
diverge and thus do not necessarily describe the exact levels, 
not even after a certain resummation (except for some important notable 
exceptions like the anharmonic oscillators with $f$ degrees of freedom 
with a polynomial perturbing potential which is asymptotically positive 
definite [29,30]), it is important to compare the 
semiclassical approximation (and the quantal perturbation results) 
with the exact spectra, which in general is impossible, since we 
generally do not have explicit solutions of the Schr\"odinger problem 
in a closed form. Therefore, we stress the importance of specific 
case studies like the present one, in order to get a better understanding 
of the quality of semiclassical mechanics. 
\par 
Finally we note that the extraction of quantum corrections 
for resonant systems is a more intricate procedure; 
some initial results for perturbed 
resonant oscillators can be found in Graffi [31].

\begin{center}
{\bf Acknowledgements}
\end{center}

\par
The author thanks Professors Sandro Graffi and Marko Robnik 
for many enlightening discussions. 

\newpage

\begin{center}
\begin{tabular}{|cccc|} \hline\hline 
  $E^{ex}$ & $E^{sc}_{n_1n_2}$ & $E^{qp}_{n_1n_2}$  & $(n_1,n_2)$\\ \hline
  1.230722 & 1.230990 & 1.230522  & (0,0) \\
  2.275974 & 2.273214 & 2.274701  & (1,0) \\
  2.689415 & 2.690856 & 2.687816  & (0,1) \\
  3.316524 & 3.308447 & 3.311808  & (2,0) \\
  3.820434 & 3.814018 & 3.812833  & (1,1) \\
  4.146646 & 4.148302 & 4.142610  & (0,2) \\
  4.354307 & 4.336609 & 4.341846  & (3,0) \\
  4.937708 & 4.915967 & 4.916677  & (2,1) \\
  5.359848 & 5.347322 & 5.345305  & (1,2) \\
  5.390110 & 5.357700 & 5.364811  & (4,0) \\
  5.603778 & 5.603248 & 5.594904  & (0,3) \\   
  6.047742 & 5.996702 & 5.999287  & (3,1) \\ 
  6.424398 & 6.371719 & 6.380706  & (5,0) \\ 
  6.546966 & 6.510986 & 6.509044  & (2,2) \\ 
  6.897049 & 6.873125 & 6.866657  & (1,3) \\ 
  7.062932 & 7.055694 & 7.044699  & (0,4) \\ 
  7.152476 & 7.056224 & 7.060684  & (4,1) \\ 
  7.457506 & 7.378668 & 7.389530  & (6,0) \\ 
  7.723943 & 7.639295 & 7.639228  & (3,2) \\ 
  8.144146 & 8.093505 & 8.088912  & (2,3) \\ 
\hline\hline
\end{tabular}
\end{center}

\vskip 0.2 truecm
{\bf Table 1}: Comparison between "exact" levels 
and levels obtained by perturbation theories. First 20 levels. 
$E^{ex}$ are "exact" levels, $E^{sc}_{n_1n_2}$ are semiclassical levels, 
and $E^{qp}_{n_1n_2}$ are levels obtained with quantum perturbation theory,
where $n_1$ and $n_2$ are the quantum number.   
$\hbar =1$, $g=10^{-1}$, $\omega_1=1$ and $\omega_2=\sqrt{2}$. 

\vskip 1.5 truecm

\begin{center}
\begin{tabular}{|ccc|} \hline\hline 
$|E^{ex}-E^{sc}_{n_1n_2}|/D$ & $|E^{ex}-E^{qp}_{n_1n_2}|/D$ & $(n_1,n_2)$ \\ 
\hline
  1.0611359$\cdot 10^{-3}$ &  1.1284242$\cdot 10^{-3}$ & (0,0) \\
  1.5578579$\cdot 10^{-2}$ &  7.1859419$\cdot 10^{-3}$ & (1,0) \\
  8.1338054$\cdot 10^{-3}$ &  9.0260478$\cdot 10^{-3}$ & (0,1)\\
  4.5591835$\cdot 10^{-2}$ &  2.6613854$\cdot 10^{-2}$ & (2,0) \\
  3.6215890$\cdot 10^{-2}$ &  4.2791300$\cdot 10^{-2}$ & (1,1) \\
  9.3476856$\cdot 10^{-3}$ &  2.2781115$\cdot 10^{-2}$ & (0,2) \\
  9.9898852$\cdot 10^{-2}$ &  7.0337765$\cdot 10^{-2}$ & (3,0) \\
  0.1227176                &  0.1187100                & (2,1) \\
  7.0703819$\cdot 10^{-2}$ &  9.2249520$\cdot 10^{-2}$ & (1,2) \\
  0.1829406                &  0.1428019                & (4,0) \\
  2.9902905$\cdot 10^{-2}$ &  5.0089385$\cdot 10^{-2}$ & (0,3) \\
  0.2880960                &  0.2735053                & (3,1) \\
  0.2973495                &  0.2466222                & (5,0) \\
  0.2030921                &  0.2140520                & (2,2) \\
  0.1350395                &  0.1715501                & (1,3) \\
  4.0854741$\cdot 10^{-2}$ &  0.1029161                & (0,4) \\
  0.5432989                &  0.5181223                & (4,1) \\
  0.4450069                &  0.3836938                & (6,0) \\
  0.4778005                &  0.4781800                & (3,2) \\
  0.2858459                &  0.3117708                & (2,3) \\
\hline\hline
\end{tabular}
\end{center}

\vskip 0.2 truecm
{\bf Table 2}: The error measured in units 
of the mean level spacing $D$ for the first 20 levels. 
$D$ is calculated for the lowest $100$ levels. 
$E^{ex}$ are "exact" levels, $E^{sc}_{n_1n_2}$ are semiclassical levels,  
and $E^{qp}_{n_1n_2}$ are levels obtained with quantum perturbation theory, 
where $n_1$ and $n_2$ are the quantum number.    
$\hbar =1$, $g=10^{-1}$, $\omega_1=1$ and $\omega_2=\sqrt{2}$. 

\newpage

\begin{center}
\begin{tabular}{|ccc|} \hline\hline 
                       $\hbar= 1$   &     $\hbar =10^{-1}$    & $(n_1,n_2)$ \\ 
\hline
           1.0611359$\cdot 10^{-3}$ &  2.4773894$\cdot 10^{-5}$ & (0,0) \\
           1.5578579$\cdot 10^{-2}$ &  1.0003044$\cdot 10^{-4}$ & (1,0) \\
           8.1338054$\cdot 10^{-3}$ &  1.8136360$\cdot 10^{-4}$ & (0,1) \\
           4.5591835$\cdot 10^{-2}$ &  2.5054353$\cdot 10^{-4}$ & (2,0) \\
           3.6215890$\cdot 10^{-2}$ &  5.6091835$\cdot 10^{-6}$ & (1,1) \\
           9.3476856$\cdot 10^{-3}$ &  3.1972348$\cdot 10^{-4}$ & (0,2) \\
           9.9898852$\cdot 10^{-2}$ &  4.3938606$\cdot 10^{-4}$ & (3,0) \\
           0.1227176                &  2.3371598$\cdot 10^{-4}$ & (2,1) \\
           7.0703819$\cdot 10^{-2}$ &  9.3486393$\cdot 10^{-5}$ & (1,2) \\
           0.1829406                &  7.0301769$\cdot 10^{-4}$ & (4,0) \\
           2.9902905$\cdot 10^{-3}$ &  4.4125578$\cdot 10^{-4}$ & (0,3) \\
           0.2880960                &  6.3570746$\cdot 10^{-4}$ & (3,1) \\
           0.2973495                &  2.3932516$\cdot 10^{-4}$ & (5,0) \\
           0.2030921                &  1.0582660$\cdot 10^{-3}$ & (2,2) \\
           0.1350395                &  1.1966258$\cdot 10^{-4}$ & (1,3) \\
           4.0854741$\cdot 10^{-2}$ &  1.2452388$\cdot 10^{-3}$ & (0,4) \\
           0.5432989                &  5.2726327$\cdot 10^{-4}$ & (4,1) \\
           0.4450069                &  8.1146188$\cdot 10^{-4}$ & (6,0) \\
           0.4778005                &  1.5294374$\cdot 10^{-3}$ & (3,2) \\
           0.2858459                &  3.0663537$\cdot 10^{-4}$ & (2,3) \\
\hline\hline
\end{tabular}
\end{center}

\vskip 0.2 truecm 
{\bf Table 3}: The error measured in units 
of the mean level spacing $D$ between "exact" levels and 
semiclassical levels. First 20 levels. 
$D$ is calculated for the lowest $100$ levels. 
$g=10^{-1}$, $\omega_1=1$ and $\omega_2=\sqrt{2}$. 

%\newpage
\vskip 1. truecm

\parindent=0.pt

\section*{References} 

[1] Ozorio de Almeida, A.M., {\it Hamiltonian Systems: Chaos and 
Quantization}, Cambridge University Press, Cambridge, 1990. 

[2] Gutzwiller, M.C., {\it Chaos in Classical and Quantum Mechanics}, 
Springer, New York, 1990.

[3] Casati, G. and Chirikov, B., {\it Quantum Chaos}, 
Cambridge University Press, Cambridge, 1990. 

[4] Maslov V.P. and Fredoriuk, M.V., {\it Semiclassical Approximation in 
Quantum Mechanics}, Reidel Publishing Company, London, 1981.  

[5] Rau, A.R.P., 'The Asymmetric Rotor as a Model for Localization', 
{\it Rev. Mod. Phys.}, {\bf 64} (1992) 623. 

[6] Braun, P.A., 'Discrete Semiclassical Methods in the Theory of Rydberg 
Atoms in External Fields', {\it Rev. Mod. Phys.}, {\bf 65} (1993) 115. 

[7] Einstein, A., 'Zum Quantensatz von Sommerfeld und Epstein', 
{\it Verhandlungen der Deutschen Physikalischen 
Gefsellschaft}, {\bf 19} (1917) 82. 

[8] Maslov, V.P., {\it Theorie des Perturbations et Methodes Asymptotiques}, 
Dunod, Paris, 1972. 

[9] Graffi, S. and Paul, T., 'The Schr\"odinger Equation and Canonical 
Perturbation Theory', {\it Commu. Math. Phys.}, {\bf 107} (1987) 25. 

[10] Degli Esposti, M., Graffi, S. and Herczynski, J., 
'Exact Quantization of the Lie Algorithm in the Bargmann Representation', 
{\it Ann. Phys.} (N.Y.), {\bf 209} (1991) 364. 

[11] Alvarez, G., Graffi S. and Silverstone, H.J., 
'Transition from Classical to Quantum Mechanics: $x^4$ Perturbed Harmonic 
Ocillator', {\it Phys. Rev.}, A {\bf 38} (1988) 1687. 

[12] Ali, M.K. and Wood, W.R., 
'The Birkhoff--Gustavson Normal Form of One--Dimensional Double Well 
Hamiltonians', {\it J. Math. Phys.}, {\bf 30} (1989) 1238. 

[13] Ali, M.K., Wood, W.R. and Devitt, S.J., 
'On the Summation of the Birkhoff--Gustavson Normal Form on an Anharmonic 
Oscillator', {\it J. Math. Phys.}, {\bf 27} (1986) 1806. 

[14] Voros, A., 'The Return of the Quartic Oscillator. 
The Complex WKB Method', 
{\it Ann. Inst. H. Poincar\`e}, {\bf A 39} (1983) 211. 

[15] Robnik, M., 'Aspects of Quantum Chaos in Generic Systems', 
to appear in {\it Proceedings of the Conference 
"Nonlinear Phenomena in Complex Systems"}, Minsk, Belarus, 
10--14 February 1996. 

[16] Pullen, R.A. and Edmonds, R.A., 
'Comparison of Classical and Quantal Spectra for a Totally Bound Potential', 
{\it J. Phys.}, {\bf A 14} (1981) L477. 

[17] Salasnich, L., 'Chaos Suppression in the SU(2) Yang--Mills--Higgs System', 
{\it Phys. Rev.}, {\bf D 52} (1995) 6189. 

[18] Salasnich, L., 'Chaos and Quantum Chaos in a Yang--Mills--Higgs System', 
{\it Mod. Phys. Lett.}, {\bf A 12} (1997) 1473.

[19] Dittrich, W. and Reuter, M., {\it Classical and Quantum Dynamics}, 
Springer, New York, 1992. 

[20] Sanders, J.A. and Verhulst, F., {\it Averaging Methods in Nonlinear 
Dynamical Systems}, Springer, Yew York, 1985. 

[21] M. Robnik, 'On the Pad\`e Approximations to the Birkhoff--Gustavson 
Normal Form', {\it J. Phys.}, {\bf A 26} (1993) 7427. 

[22] Messiah, A., {\it Mecanique Quantique}, Dunod, Paris, 1962. 

[23] Graffi, S., Manfredi, V.R. and Salasnich, L., 
'Accuracy of the Semiclassical Approximation: the Pullen--Edmonds 
Hamiltonian', {\it Nuovo Cim.}, {\bf B 109} (1994) 1147. 

[24] Robnik, M. and Salasnich, L., 
'WKB to All Orders and the Accuracy of the Semiclassical Approximation', 
{\it J. Phys.}, {\bf A 30} (1997) 1711. 

[25] Robnik, M. and Salasnich, L., 
'WKB Expansion for the Angular Momentum and the Kepler Problem: 
from the Torus Quantization to the Exact One', 
{\it J. Phys.}, {\bf A 30} (1997) 1719. 

[26] Salasnich, L. and Sattin, F., 'SWKB for the Angular Momentum', 
{\it Mod. Phy. Lett.}, {\bf B 11} (1997) 801.

[27] Salasnich, L. and Sattin, F., 
'On the Convergence of the WKB series for the Angular Momentum Operator', 
{\it J. Phys.}, {\bf A 30} (1997) 7597. 

[28] Bender, C. and Orzag, S., {\it Advanced Mathematical Methods for 
Scientists and Engeneers}, Mc Graw Hill, Yew York, 1978.

[29] Graffi, S., Grecchi, V. and Simon, B., 
'Borel Summability: Application to the Anharmonic Oscillator', 
{\it Phys. Lett.}, {\bf B 32} (1970) 631. 

[30] Reed, M. and Simon, B., {\it Methods of Modern Mathematical Physics}, 
Academic Press, New York, 1978.

[31] Graffi, S., 'Exact Quantization of Canonical Perturbation Theory', 
In: Guerra, F., Loffredo, M.I. and Marchioro, C. (eds), 
{\it Probabilistic Methods in Mathematical Physics}, 
World Scientific, Singapore, 1992.

\end{document}